\documentclass[3p]{elsarticle}

\usepackage{lineno,hyperref}
\usepackage{amsmath,amssymb}
\modulolinenumbers[5]
\usepackage{xcolor}

\usepackage{comment}

\newdefinition{df}{Definition}
\newdefinition{rmk}{Remark}
\newtheorem{thm}{Theorem}
\newproof{pf}{Proof}

\newcommand{\bfv}{\boldsymbol{v}}
\newcommand{\bfx}{{\boldsymbol{x}}}
\newcommand{\bfu}{\boldsymbol{u}}

\allowdisplaybreaks

\begin{document}

\begin{frontmatter}

\title{Maximum entropy principle approach to a non-isothermal\\ Maxwell-Stefan diffusion model} \tnotetext[mytitlenote]{Authors acknowledge financial support (B.A.) from the COST Action, CA18232 MAT-DYN-NET, supported by the ``European Cooperation in Science and Technology'' through the ``Short
Term Scientific Mission'' at the Department of Mathematics and Informatics, University of Novi Sad, Serbia, and (S.S.) from the Ministry of Education, Science and Technological Development of the Republic of Serbia (Grant No. 451-03-9/2021-14/200125).}


\author[BAaddress,BAaddressb]{Benjamin Anwasia}
\address[BAaddress]{Universidade do Minho, Centro de Matem\'atica, Campus de Gualtar, 4710-057 Braga, Portugal}
\address[BAaddressb]{New York University Abu Dhabi, Abu Dhabi, United Arab Emirates}
\ead{id6226@alunos.uminho.pt}

\author[SSaddress]{Srboljub Simi\'{c}\corref{mycorrespondingauthor}}
\address[SSaddress]{Department of Mathematics and Informatics, Faculty of Sciences, University of Novi Sad, Trg Dositeja Obradovi\'{c}a 3, 21000 Novi Sad, Serbia}
\cortext[mycorrespondingauthor]{Corresponding author}
\ead{ssimic@uns.ac.rs}

%\address[BAaddress]{1600 John F Kennedy Boulevard, Philadelphia}
%\address[SSaddress]{360 Park Avenue South, New York}

\begin{abstract}
In this study we apply the maximum entropy principle to derive the properly scaled velocity distribution function of Boltzmann equations for mixtures, which leads to a non-isothermal Maxwell-Stefan diffusion model. We also analyze the entropy balance law and derive the kinetic  entropy production from the scaled distribution function. 
\end{abstract}

\begin{keyword}
diffusion \sep Maxwell-Stefan model \sep maximum entropy principle 
\MSC[2010] 35Q20 \sep 82C40 \sep 76R50
\end{keyword}

\end{frontmatter}

\linenumbers

\section{Introduction} 

Diffusion is a process of motion of one specie relative to another. If the motion occurs relative to a background medium, the process is usually described by the Fick's law or some of its generalizations. When cross-diffusion is more prominent, like in gaseous mixtures, the Maxwell-Stefan diffusion model is more appropriate \cite{krishna1997maxwell}. It turned out that cross-diffusion models of Maxwell-Stefan kind can be used as a starting point for diverse applications, ranging from engineering \cite{krishna2016describing} to medicine \cite{boudin2010diffusion}, and they also inspired new approaches to diffusion phenomena \cite{hirschler2016maxwell}.
%,chai2019maxwell,francuu2020alternative

Along with applications, mathematical analysis of Maxwell-Stefan equations emerged as equally important but relatively new subject \cite{bothe2011maxwell,herberg2017reaction}. It covers the aspects of existence and uniqueness of solutions \cite{chen2015analysis,hutridurga2018existence}, large-time asymptotics \cite{daus2020exponential} and relaxation limits \cite{salvarani2018relaxation}. Our concern, however, will be the formal derivation of Maxwell-Stefan diffusion equations and analysis of their dissipative character. 
%, as well as numerical procedures which may be sensitive on the structure of diffusion matrix $k_{ij}$ [???]

From the macroscopic point of view, Maxwell-Stefan diffusion model consists of mass balance equations for the constituents, and a kind of momentum balance relations adjoined with a closure relation due to their linear dependence. It can be derived either in the continuum framework \cite{giovangigli1999multicomponent,bothe2015continuum}, or starting from the Boltzmann equations (BEs) for mixtures \cite{boudin2015maxwell,boudin2017maxwell,anwasia2020formal,anwasia2020maxwell}. In the latter approach, crucial step in derivation of Maxwell-Stefan equations is the asymptotic limit enabled through appropriate scaling of the governing equations.

In this study we shall take the kinetic approach to Maxwell-Stefan equations. It is based upon \emph{formal} diffusive scaling of the BEs, and crucial \emph{assumption} that velocity distribution functions have the following form: 
\begin{equation} \label{Intro:f_i}
  f^\alpha_{i}(t,{\bfx},{\bfv}) = \rho_{i}^\alpha \left( \frac{m_{i}}{2 \pi k_{\mathrm{B}} T} \right)^{3/2} 
    \exp \left( - \frac{m_{i} |{\bfv} - \alpha {\bfu}^\alpha_{i}|^{2}}{2 k_\mathrm{B} T}\right), 
    \quad i = 1, \ldots, S,
\end{equation}
for some small parameter $\alpha \ll 1$, where ${\bfv}\!\in\!\mathbb{R}^{3}$ are the particle velocities, ${\bfu}^\alpha_{i} \!\in\!\mathbb{R}^{3}$ are the macroscopic species velocities, $m_{i}$ are the species atomic masses, $k_{\mathrm{B}}$ is the Boltzmann constant, $\rho_{i}^\alpha$ are the species mass densities and $T$ is the mixture temperature. 

Instead of assuming, our aim is to \emph{derive} the scaled velocity distribution function \eqref{Intro:f_i} appropriate for description of diffusion processes. To that end, we shall apply the maximum entropy principle (MEP) in the scaled (dimensionless) form (Theorem \ref{TH:Vdf}). Moreover, we shall put our analysis in a broader context of non-isothermal processes. We shall also derive the entropy balance law in the scaled form and compute the entropy production (Theorem \ref{TH:Entropy_production}) for the non-isothermal Maxwell-Stefan model, starting from the scaled velocity distribution function. Computation of the velocity distribution function from MEP will be \emph{exact}. Therefore, we shall derive the moment equations (Maxwell-Stefan model) and the entropy balance law for the mixture of monatomic Maxwellian particles, and in such a way retain the exact character of the results. 

%The paper is organized as follows. First we give a brief account on the Boltzmann equations for mixtures and their scaled version. Then we formulate the MEP and derive the properly scaled velocity distribution function. Finally, we recall the macroscopic equations for moments, which in asymptotic limit recover the non-isothermal Maxwell-Stefan diffusion equations, and derive the entropy production. 

%%%%%% 

\section{The Boltzmann equation for inert mixtures}

%%%%%%
\begin{comment}
The Boltzmann equation is the model that governs the evolution of ideal gases in kinetic theory. %In the kinetic theory of gases, the state of a gas is, modelled by a distribution function. 
It describes the evolution of a distribution function using the fact that the distribution function changes in time due to the free transport of particles and collisions between the particles that composes the gas. %as described by the Boltzmann equation.
\end{comment}
%%%%%%%%%%%%%%%%%%%%%%%%%%%%%%%%%%%%%%%%%%%%%%%%%%%%%%%%%%%%%%%%%%%%%%
We consider a mixture of monatomic ideal inert gases: a gaseous mixture of non-reactive species consisted of particles endowed with translational degrees of freedom only. For each species ${\cal A}_i$ with $i=1,2,3\dots,S$, let $f_i:=f_i(t,{\bfx}, {\bfv})$ be its distribution function, representing the densities of particles of species ${\cal A}_i$ which at time $t\in\mathbb{R}^+$ are located at position ${\bfx}\in \mathbb{R}^3$ and have velocity ${\bfv}\in \mathbb{R}^3$. In the absence of external forces, the evolution of the non-reactive mixture can be described by the following system of BEs:
\begin{equation}
\begin{aligned}
\frac{\partial f_i}{\partial t}+{\bfv}\cdot \frac{\partial f_i }{\partial {\bf{x}}}
&=\underbrace{\int_{\mathbb{R}^3}\int_{\mathbb{S}^2}\Big[f_i({\bfv}')f_i({\bfv}_*')-f_i({\bfv})f_i({\bfv}_*)\Big]\,B_i({\bfv},{\bfv}_*, \boldsymbol{\sigma})\, d\boldsymbol{\sigma}\,d{\bfv}_*}_{Q_i^{m}(f_i, f_i)}
\\[-0.4cm]
&+\sum_{\substack{j=1\\ j\neq i}}^{S}\underbrace{\int_{\mathbb{R}^3}\int_{\mathbb{S}^2}\Big[f_i({\bfv}')f_j({\bfv}_*')-f_i({\bfv})f_j({\bfv}_*)\Big]\,B_{ij}({\bfv},{\bfv}_*, \boldsymbol{\sigma})\, d\boldsymbol{\sigma}\,d{\bfv}_*}_{Q^b_{ij}(f_i, f_j)},  
\label{eq:system_of_Boltzmann_equations}
\end{aligned}
\end{equation}
for $i=1,2,3,\dots, S,$ where the first and second terms on the right-hand side describe collisions between particles of the same species and between particles of different species, respectively. They are called the mono-species and bi-species collision operators
%\begin{align}
%	Q_i^{m}(f_i, f_i) &= \int_{\mathbb{R}^3} \int_{\mathbb{S}^2}\Big[f_i(t,{\bfx},{\bfv}')f_i(t,{\bfx},{\bfv}_*')-f_i(t,{\bfx},{\bfv})f_i(t,{\bfx},{\bfv}_*)\Big]\,B_i({\bfv},{\bfv}_*, \boldsymbol{\sigma})\, d\boldsymbol{\sigma}\,d{\bfv}_*,
%	\label{eq:mono_species_collision_operator}
%	\\
%	Q^b_{ij}(f_i, f_j)&=\int_{\mathbb{R}^3}\int_{\mathbb{S}^2}\Big[f_i(t,{\bfx},{\bfv}')f_j(t,{\bfx},{\bfv}_*')-f_i(t,{\bfx},{\bfv})f_j(t,{\bfx},{\bfv}_*)\Big]\,B_{ij}({\bfv},{\bfv}_*, \boldsymbol{\sigma})\, d\boldsymbol{\sigma}\,d{\bfv}_*,
%	\label{eq:bi_species_collision_operator}
%\end{align}
with ${\bfv}'$, ${\bfv}'_*$ defined %in terms of ${\bfv}$, ${\bfv}_*$ and $\boldsymbol{\omega}$ 
as follows:
\begin{equation}
	{\bfv}'=\frac{1}{m_i+m_j}\big(m_i{\bfv}+m_j{\bfv}_*+m_j|{\bfv}-{\bfv}_*|\boldsymbol{\sigma}), \qquad
	{\bfv}_*'=\frac{1}{m_i+m_j}\big(m_i{\bfv}+m_j{\bfv}_*-m_i|{\bfv}-{\bfv}_*|\boldsymbol{\sigma}),
	\label{eq:post_collision_velocities}
\end{equation}
where $\boldsymbol{\sigma}\in\mathbb{S}^2$ is an arbitrary unit vector. The formulas given in equation \eqref{eq:post_collision_velocities} represent all the possible solutions $({\bfv}',{\bfv}'_*)\in \mathbb{R}^3 \times\mathbb{R}^3$ of the system of equations composed of the momentum and kinetic energy conservation law, which are respectively defined as follows:
\begin{equation}
	m_i{\bfv}+m_j{\bfv}_{*} =m_i{\bfv}'+m_j{\bfv}'_{*},
	\qquad
	\frac{1}{2}m_i|{\bfv}|^2+\frac{1}{2}m_j|{\bfv}_{*}|^2=\frac{1}{2}m_i|{\bfv}'|^2+\frac{1}{2}m_j|{\bfv}'_{*}|^2.
	\label{eq:conservation_of_momentum_and_kinetic_energy}
\end{equation}
%%%%%%%%%%%%%%%%%%%%%%%%%%%%%%%%%%%%%%%%%%%%%%%%%%%%%%%%%%%%%%%%%%%%%

\subsection{Scaled system of Boltzmann equations}

To write the system \eqref{eq:system_of_Boltzmann_equations} of BEs in dimensionless form, we first choose macroscopic scales for the independent variables $t, {\bfx}$ and ${\bfv}$ associated with the distribution function.To this end, let $\tau, L$ and $T_0$ respectively be the macroscopic time, space and reference temperature scales. These macroscopic scales define two macroscopic velocity scales: ${u}_0=\frac{L}{\tau}$ and ${c}_0=\sqrt{\frac{5k_B T_0}{3m_0}}$,
where ${u}_0$ is the speed at which some macroscopic portion of the gas is transported over a distance $L$ in time $\tau$, while ${c}_0$ is the speed of sound in a monatomic gas with $m_0$ the average atomic mass of the mixture. Using the macroscopic scales: $\tau, L$ and ${c}_0$, one can define the dimensionless time, space, velocity, distribution function and cross-sections:% respectively, as follows:
\begin{align}
	&\displaystyle\widehat{t}=\frac{t}{\tau},\quad  \widehat{\bfx}=\frac{\bfx}{L},\qquad \widehat{\bfv} =\frac{{\bfv}}{c_0},\quad {\widehat{f}_i}(\widehat{t},\widehat{\bfx},\widehat{\bfv})=\frac{L^3c_0^3}{N}f_i(t,{\bfx},{\bfv}),\quad\widehat{B}_i(\widehat{\bfv},\widehat{\bfv}_*,\widehat{\boldsymbol{\sigma}})=\frac{1}{c_0 4\pi r^2}B_i({\bfv},{\bfv}_*,\boldsymbol{\sigma}),
	\label{eq:dimensionless_variables_distribution_function_and_collision_cross-sections}
\end{align}
respectively,
and rewrite the system \eqref{eq:system_of_Boltzmann_equations} of BEs in dimensionless form in which only two parameters appear---Mach number $\mathrm{Ma}$ which multiplies $\partial \widehat{f_i}/\partial \widehat{t}$, and a reciprocal of the Knudsen number $\mathrm{Kn}$ which multiplies the scaled collision operators:
\begin{equation*}
  \mathrm{Ma} = \frac{u_0}{c_0}, \quad 
  \mathrm{Kn} = \frac{\text{mean free path}}{\text{macroscopic length scale}}
    %= \left( \frac{L^3}{N\times 4\pi r^2} \right) \div L
    = \frac{L^3}{N\times 4\pi r^2}\times \frac{1}{L}.
\end{equation*}
In the sequel we shall assume that $\mathrm{Ma} = \mathrm{Kn} = \alpha \ll 1$, which amounts to diffusive scaling of the BEs \eqref{eq:system_of_Boltzmann_equations}. Removing the hats for simplicity and adding superscript $\alpha$ to the distribution function so as to differentiate the dimensionless system of BEs from equation \eqref{eq:system_of_Boltzmann_equations}, we obtain:
%defining the Knudsen and Mach numbers $(Kn)$, $Ma$, respectively by
%and then removing the hats for simplicity as well as adding a superscript $\alpha$ to the distribution function so as to differentiate the dimensionless system of Boltzmann equations from equation \eqref{eq:system_of_Boltzmann_equations}, we obtain that  the dimensionless Boltzmann equation is given as follows:
%\begin{equation}
%	Ma\frac{\partial \widehat{f_i}}{\partial \widehat{t}}+\widehat{\bfv}\cdot\frac{\partial \widehat{f_i}}{\partial \widehat{\bfx}}=\frac{1}{Kn} \widehat{Q}_i^{m}(\widehat{f}_i, \widehat{f}_i)+\frac{1}{Kn}  \sum_{\substack{j=1\\ j\neq i}}^{S}\widehat{Q}^b_{ij}(\widehat{f}_i, \widehat{f}_j),
%	\label{eq:dimensionless_system_of_Boltzmann_equations1}   
%\end{equation}
%where $Ma=u_0/c_0$ is a dimensionless parameter called the Mach number $Ma$. If one assumes that:
%and then removing the hats for simplicity as well as adding a superscript $\alpha$ to the distribution function so as to differentiate the dimensionless system of Boltzmann equations from equation \eqref{eq:system_of_Boltzmann_equations}, we obtain:
\begin{equation}
	\alpha\frac{\partial f_i^\alpha}{\partial t}+{\bfv}\cdot\frac{\partial f_i^\alpha}{\partial {\bfx}}=\frac{1}{\alpha} {Q}_i^{m}(f^\alpha_i, f^\alpha_i)+\frac{1}{\alpha}  \sum_{\substack{j=1\\ j\neq i}}^{S}{Q}^b_{ij}(f^\alpha_i, f^\alpha_j).
	\label{eq:dimensionless_system_of_Boltzmann_equations_final} 
\end{equation}
%where we have assumed that $Ma=Kn=\alpha \ll 1$, with $Ma$, $Kn$ the Mach and Knudsen numbers, respectively defined as follows:
%\begin{equation*}
%	Kn=\frac{\text{mean free path}}{\text{macroscopic length scale}}=\bigg(\frac{L^3}{N\times 4\pi r^2}\bigg)\div L=\frac{L^3}{N\times 4\pi r^2}\times \frac{1}{L}, \qquad Ma=\frac{u_0}{c_0},
%\end{equation*}
%where the collision operators ${Q}_i^{m}(f^\alpha_i, f^\alpha_i)$ and ${Q}^b_{ij}(f^\alpha_i, f^\alpha_j)$ formally have the same form as in {\color{blue}\eqref{eq:system_of_Boltzmann_equations}}.
%\begin{align}
%	Q_i^{m}(f^\alpha_i, f^\alpha_i) &= \int_{\mathbb{R}^3} \int_{\mathbb{S}^2}\Big[f_i^\alpha(t,{\bfx},{\bfv}')f_i^\alpha(t,{\bfx},{\bfv}_*')-f_i^\alpha(t,{\bfx},{\bfv})f_i^\alpha(t,{\bfx},{\bfv}_*)\Big]\,B_i({\bfv},{\bfv}_*, \boldsymbol{\sigma})\, d\boldsymbol{\sigma}\,d{\bfv}_*,
%	\label{eq:scaled_mono_species_collision_operator}
%	\\
	%\sum_{\substack{j=1\\ j\neq i}}^{S}
%	Q^b_{ij}(f^\alpha_i, f^\alpha_j)&=
	%\sum_{\substack{j=1\\ j\neq i}}^{S}
%	\int_{\mathbb{R}^3}\int_{\mathbb{S}^2}\Big[f_i^\alpha(t,{\bfx},{\bfv}')f_j^\alpha(t,{\bfx},{\bfv}_*')-f_i^\alpha(t,{\bfx},{\bfv})f_j^\alpha(t,{\bfx},{\bfv}_*)\Big]\,B_{ij}({\bfv},{\bfv}_*, \boldsymbol{\sigma})\, d\boldsymbol{\sigma}\,d{\bfv}_*.
%	\label{eq:scaled_bi_species_collision_operator}
%\end{align}
%%%%%%%%%%%%%%%%%%%%%%%%%%%%%%%%%%%%%%%%%%%%%%%%%%%%%%%%%%%%%%%%%%%%%%
\subsection{Moment Equations}
Multiplying the scaled system of BEs \eqref{eq:dimensionless_system_of_Boltzmann_equations_final} by an arbitrary function of the velocity variable $\psi_i({\bfv})$, %$m_i$, $m_i{\bfv}$ and $m_i|{\bfv}|^2$, respectively, 
and then integrate the resulting equation over all values of the velocity ${\bfv}$ gives the moment equations:
\begin{equation}
\alpha\frac{\partial}{\partial t}\int_{\mathbb{R}^3}\!\!   \psi_i({\bfv})f_i^\alpha \,d{\bfv}+\frac{\partial}{\partial {\bfx}}\int_{\mathbb{R}^3}\!\! {\bfv}  \psi_i({\bfv}) f_i^\alpha \,d{\bfv}
\!=\!\frac{1}{\alpha} \int_{\mathbb{R}^3}\!\!  \psi_i({\bfv}) {Q}_i^{m}(f^\alpha_i, f^\alpha_i)\,d{\bfv}+\frac{1}{\alpha}  \sum_{\substack{j=1\\ j\neq i}}^{S}\int_{\mathbb{R}^3}\!\!  \psi_i({\bfv}){Q}^b_{ij}(f^\alpha_i, f^\alpha_j)\,d{\bfv},
	\label{eq:moment_equations} 
\end{equation}
where the integrals on the right-hand side are the weak formulation of the mono-species and bi-species collision operators respectively defined as follows:
\begin{align}
	\nonumber&\int_{\mathbb{R}^3} \psi_{i}({\bfv})
	Q_i^{m}(f^\alpha_i, f^\alpha_i)\,d{\bfv} =-\frac{1}{4} \int_{\mathbb{R}^3}\int_{\mathbb{R}^3} \int_{\mathbb{S}^2}\Big[f_i^\alpha({\bfv}')f_i^\alpha({\bfv}_*')-f_i^\alpha({\bfv})f_i^\alpha({\bfv}_*)\Big]
	\\ 
	&\hspace{6.1cm}
	\times\big[\psi_i({\bfv}')+\psi_i({\bfv}'_*)-\psi_i({\bfv})-\psi_i({\bfv}_*)\big]B_i({\bfv},{\bfv}_*, \boldsymbol{\sigma})\, d\boldsymbol{\sigma}\,d{\bfv}_*\,d{\bfv},
	\label{eq:weak_form_mono_species_collision_operator}
	\\
&\int_{\mathbb{R}^3} \psi_{i}({\bfv})
	Q^b_{ij}(f^\alpha_i, f^\alpha_j)\,d{\bfv}\!=\!
	\int_{\mathbb{R}^3}\int_{\mathbb{R}^3}\int_{\mathbb{S}^2} f_i^\alpha({\bfv})f_j^\alpha({\bfv}_*) \Big[\psi_i({\bfv}')-\psi_i({\bfv}) \Big] B_{ij}({\bfv},{\bfv}_*, \boldsymbol{\sigma})\, d\boldsymbol{\sigma}\,d{\bfv}_*d{\bfv}.
	\label{eq:weak_form_bi_species_collision_operator}
\end{align}

%%%%%% 

\section{Maximum entropy principle} 

%%%%%% 

\begin{comment}
The maximum entropy principle (MEP) has its {\color{blue}physical motivation} in the fact that equilibrium (Maxwellian) distribution function maximizes the kinetic entropy. This result may be expressed as a solution of the variational problem with constraints, in which the kinetic entropy is the objective function, while its moments are taken as constraints. A major breakthrough {\color{blue}in the context of kinetic theory} came through the work of Kogan \cite{kogan1969rarefied} who found out that Grad's velocity distribution function shares the same property. These results were expanded and widely used in the extended thermodynamics [ET1ed,Dreyer], while their strict mathematical framework was developed later in [Levermore]. Recently, this approach was further generalized to the analysis of polyatomic gases [MEP]. 
%Such an interpretation is deeply physically motivated: the state of the system at mesoscopic level is described by means of velocity distribution function, whereas the information upon it is collected through its moments, i.e. macroscopic field variables. 
%
%although in somewhat different sense (exact minimizer had to be expanded in the neighborhood of Maxwellian due to divergence of the moments). 

Our aim is to apply MEP and \emph{derive} the properly scaled velocity distribution function which leads to a non-isothermal Maxwell-Stefan model. To that end we introduce the following
\end{comment}
The maximum entropy principle (MEP) determines the approximate velocity distribution function, compatible with selected moments, through maximization of the kinetic entropy \cite{kogan1969rarefied,muller1993extended,dreyer1987maximisation,levermore1996moment,pavic2013maximum}. Our aim is to apply MEP and \emph{derive} the properly scaled velocity distribution function which leads to a non-isothermal Maxwell-Stefan model. To that end we introduce the following
\begin{df}
Kinetic entropy $H(t,{\bfx})$ is defined as:
\begin{equation} \label{MEP:Entropy}
  H(t,{\bfx}) := \sum_{i=1}^{S} H_{i}(t,{\bfx}), \quad
  H_{i}(t,{\bfx}) := - k_{\mathrm{B}} \int_{\mathbb{R}^{3}} f_{i} \log (b_{i} f_{i}) d{\bfv},
\end{equation}
where $b_{i}$ is the dimensional constant which makes the argument of the $\log$ function dimensionless. Moments of the distribution functions, representing partial mass, momentum and energy densities, are defined as: 
\begin{equation} \label{MEP:Moments}
\begin{split}
  \rho_{i}(t,{\bfx}) := \int_{\mathbb{R}^{3}} m_{i} f_{i} d{\bfv}, & \quad 
  \rho_{i}(t,{\bfx}) {\bfu}_i(t,{\bfx}) := 
    \int_{\mathbb{R}^{3}} m_{i} {\bfv} f_{i} d{\bfv}, 
    \\
  \rho_{i}(t,{\bfx}) |{\bfu}_i(t,{\bfx})|^{2} 
    + 3 & \rho_{i}(t,{\bfx}) \frac{k_{B}}{m_{i}} T(t,{\bfx}) 
    := \int_{\mathbb{R}^{3}} m_{i} |{\bfv}|^{2} f_{i} d{\bfv}, 
\end{split}
\end{equation}
where $\rho_{i}$ is the mass density and ${\bfu}_i$ is the mean velocity of the specie $\mathcal{A}_{i}$, while $T$ is the common kinetic temperature.
\end{df} 

%\begin{rmk}
It will be assumed that all the species have the same kinetic temperature $T$. This simplifying assumption is appropriate for non-isothermal diffusion problems, see \cite{hutridurga2017maxwell}. 
%\end{rmk}

In the sequel we shall use the dimensionless macroscopic variables: 
\begin{equation}
  \widehat{\rho}_{i} = \frac{L^{3}}{m_{i} N} \rho_{i}, \quad
  \widehat{\bfu}_{i} = \frac{\bfu_{i}}{u_{0}}, \quad 
  \widehat{T} = \frac{T}{T_{0}}, \quad 
%  \widehat{\varepsilon}_{i} = \frac{\varepsilon_{i}}{c_{0}^{2}} 
%    = \frac{3 m_{0}}{5 m_{i}} \frac{3}{2} \widehat{T}, \quad 
  \widehat{H}_{i} = \frac{L^{3}}{k_{\mathrm{B}} N} H_{i}, \quad 
  \widehat{b}_{i} = \frac{N}{L^{3} c_{0}^{3}} b_{i}.
  \label{eq:dimensionless_macroscopic_variables}
\end{equation}
and drop the hat notation for convenience. The variational problem (MEP) may now be formulated in dimensionless form as follows. 

\paragraph{Maximum entropy principle} Find the velocity distribution functions $f_{i}^{\alpha}(t,{\bfx},{\bfv})$ which maximize the kinetic entropy: 
\begin{equation} \label{MEP:Entropy_DL}
  H^{\alpha}(t,{\bfx}) = - \int_{\mathbb{R}^{3}} \sum_{i=1}^{n} f_{i}^{\alpha} \log (b_{i} f_{i}^{\alpha}) 
    d{\bfv} \to \mathrm{max}, 
\end{equation}
subject to the following constraints:
\begin{equation} \label{MEP:Moments_DL}
\begin{split}
  \rho_{i}^{\alpha}(t,{\bfx}) 
    = \int_{\mathbb{R}^{3}} f_{i}^{\alpha} d{\bfv}, & \quad 
	\alpha \rho_{i}^{\alpha}(t,{\bfx}) 
	{\bfu}_i^{\alpha}(t,{\bfx}) = 
	\int_{\mathbb{R}^{3}} {\bfv} f_{i}^{\alpha} d{\bfv}, 
  \\
  \alpha^{2} \rho_{i}^{\alpha}(t,{\bfx}) 
    |{\bfu}_i^{\alpha}(t,{\bfx})|^{2} 
    + & 3 \rho_{i}^{\alpha}(t,{\bfx}) \frac{3}{5} \frac{m_{0}}{m_{i}} T(t,{\bfx}) 
	= \int_{\mathbb{R}^{3}} |{\bfv}|^{2} f_{i}^{\alpha} d{\bfv}.  
\end{split}
\end{equation}

We would like to note that small parameter $\alpha$ naturally appears in dimensionless constraints \eqref{MEP:Moments_DL} due to scaling introduced in previous Section. 

\begin{thm} \label{TH:Vdf}
The velocity distribution functions which solve the variational problem \eqref{MEP:Entropy_DL}-\eqref{MEP:Moments_DL} have the following form: 
\begin{equation}
  f_{i}^{\alpha}(t,{\bfx},{\bfv}) = \left( \frac{5}{3 m_{0}} \right)^{3/2} 
    \rho_{i}^{\alpha}(t,{\bfx}) 
    \left( \frac{m_{i}}{2 \pi T(t,{\bfx})} \right)^{3/2} 
    \exp \left( - \frac{5}{3 m_{0}} 
    \frac{m_{i} |{\bfv} - \alpha {\bfu}_i^{\alpha}(t,{\bfx})|^{2}}{2 T(t,{\bfx})} \right).
    \label{eq:scaled_distribution_function}
\end{equation}
\end{thm}

\begin{pf}
Proof of the Theorem \ref{TH:Vdf} relies on the method of multipliers. Starting from \eqref{MEP:Entropy_DL}-\eqref{MEP:Moments_DL} we define extended functional: 
\begin{equation*}
  \mathcal{H}^{\alpha}(t,{\bfx}) 
    := \int_{\mathbb{R}^{3}} \sum_{i=1}^{n} \left( f_{i}^{\alpha} 
    \log (b_{i} f_{i}^{\alpha}) + \lambda_{i}^{(0)} f_{i}^{\alpha} 
    + \boldsymbol{\lambda}_{i}^{(1)} \cdot {\bfv} f_{i}^{\alpha} 
    + \lambda_{i}^{(2)} |{\bfv}|^{2} f_{i}^{\alpha} \right) d{\bfv}
    \equiv \int_{\mathbb{R}^{3}} \mathcal{L} ({\bfv}, f_{i}^{\alpha}, 
    \nabla_{{\bfv}}f_{i}^{\alpha}) d{\bfv}, 
\end{equation*}
where $\lambda_{i}^{(0)}(t,{\bfx}) \in \mathbb{R}$, $\boldsymbol{\lambda}_{i}^{(1)}(t,{\bfx}) \in \mathbb{R}^{3}$ and $\lambda_{i}^{(2)}(t,{\bfx}) \in \mathbb{R}$ are the unknown multipliers. Since the density function $\mathcal{L}$ does not depend on $\nabla_{{\bfv}}f_{i}^{\alpha}$, necessary condition for extremum is reduced to $\partial \mathcal{L}/\partial f_{i}^{\alpha} = 0$, $i = 1, \ldots, S$, which yields $S$ uncoupled equations that can be solved for $f_{i}^{\alpha}$: 
\begin{equation*}
  f_{i}^{\alpha} = \frac{1}{b_{i}} \exp \left[ - \left( 1 + \lambda_{i}^{(0)} 
    + \boldsymbol{\lambda}_{i}^{(1)} \cdot {\bfv} + \lambda_{i}^{(2)} |{\bfv}|^{2} \right) \right].
\end{equation*} 
Plugging the last expression into the constraints \eqref{MEP:Moments_DL}, after rather straightforward computation the following relations which determine the multipliers are obtained: 
\begin{equation*}
  \lambda_{i}^{(2)} = \frac{5}{3} \frac{m_{i}}{m_{0}} \frac{1}{2 T}, \quad 
	\frac{\boldsymbol{\lambda}_{i}^{(1)}}{2 \lambda_{i}^{(2)}} 
	= - \alpha {\bfu}_i^{\alpha}, \quad 
    \frac{1}{b_{i}} \exp \left(- 1 - \lambda_{i}^{(0)} \right) 
	\exp \left( \frac{|\boldsymbol{\lambda}_{i}^{(1)}|^{2}}{4 \lambda_{i}^{(2)}} \right) 
	\left( \frac{\pi}{\lambda_{i}^{(2)}} \right)^{3/2} = \rho_{i}^{\alpha}.
\end{equation*}
The proof is completed by returning these expressions into the maximizer. 
\end{pf}

\begin{comment}
\begin{rmk}
Note that maximizer \eqref{eq:scaled_distribution_function} is exact: the moments \eqref{MEP:Moments_DL} computed using \eqref{eq:scaled_distribution_function} are convergent and maximizer does not require expansion.  
\end{rmk}
\end{comment}

%%%%%% 

%\section{Non-isothermal Maxwell-Stefan diffusion equations and the entropy balance law} 

%%%%%% 

%The non-isothermal diffusion system of equations where the diffusion process is described by the Maxwell-Stefan equation is obtained from the energy conservation equation for the mixture, the concentration and momentum balance equations for the species. On the other hand, the entropy balance is obtained from the entropy density, flux and production.

%%%%%%%%%%%%%%%%%%%%%%%%%%%%%%%%%%%%%%%%%%%%%%%%%%%%%%%%%%%%%%%%%%%%%%%

\section{Non-isothermal Maxwell-Stefan diffusion equations} 

The Maxwell-Stefan diffusion equations for the mixture of Maxwell molecules were derived in \cite{boudin2015maxwell}, and in the non-isothermal setting in \cite{hutridurga2017maxwell}, while in \cite{anwasia2020maxwell} the case of reactive mixture of polyatomic gases was considered. Since our interest is on inert mixtures, we shall give the non-isothermal  Maxwell-Stefan diffusion equations here for the completeness. They can be obtained from the moment equations \eqref{eq:moment_equations} by using the scaled distribution function \eqref{eq:scaled_distribution_function} obtained by MEP. The mass/concentrantion  and momentum balance laws for the species are obtained taking $\psi_{i}(\bfv) = 1$ and $\psi_{i}(\bfv) = \bfv$, respectively. The energy conservation law for the mixture is obtained taking $\psi_{i}(\bfv) = |\bfv|^{2}/2$ and summing up all the energy balance laws for the species. The resulting equations read:  
\begin{equation}
\begin{aligned}
& \frac{\partial \rho_{i}^\alpha}{\partial t} 
  + \frac{\partial }{\partial {\bfx}}( \rho_{i}^\alpha {\bfu}_i^\alpha) = 0, 
\\
& \alpha^2 \left[ \frac{\partial }{\partial t}(\rho_{i}^\alpha {\bfu}_i^{\alpha}) 
  + \frac{\partial }{\partial {\bfx}} \left( \rho_{i}^\alpha{\bfu}_i^\alpha \otimes {\bfu}_i^\alpha \right) \right] + 
	\frac{3 m_{0}}{5 m_i} \frac{\partial}{\partial {\bfx}}(\rho_{i}^\alpha T) 
	= 
		%\int_{\mathbb{R}^3}\!\! \bfv {Q}_i^{m}(f^\alpha_i, f^\alpha_i)\,d{\bfv} 
	2\pi\sum_{\substack{j=1\\ j\neq i}}^{S}||b_{ij}||_{L^1} \frac{m_j}{m_i+m_j}\rho_i^{\alpha} \rho_j^{\alpha} ({\bfu}^{\alpha}_j-  {\bfu}^{\alpha}_i)
%		\sum_{\substack{j=1\\ j\neq i}}^{S}\frac{m_j}{m_i+m_j}2\pi||b_{ij}||_{L^1}\big(c^\alpha_i c^\alpha_j{\bfu}^\alpha_j-c^\alpha_i c^\alpha_j{\bfu}^\alpha_i\big),}
\\
& \frac{\partial }{\partial t}\sum_{i=1}^{S} \left( 
  \alpha^2 \frac{1}{2} \rho_{i}^\alpha |{\bfu}_i^{\alpha}|^2 + \frac{3 m_{0}}{5 m_i} \frac{3}{2} \rho_{i}^\alpha T \right) 
  + \frac{\partial }{\partial {\bfx}} \sum_{i=1}^{S} \left[ \left( \alpha^2 \frac{1}{2} \rho_{i}^\alpha |{\bfu}_i^{\alpha}|^2 + \frac{3 m_{0}}{5 m_i} \frac{5}{2} \rho_{i}^\alpha T \right) 
  {\bfu}_i^{\alpha} \right] 
  = 0.
\end{aligned}
\label{eq:Macroscopic_equations}
\end{equation}
Taking the limit as $\alpha \to 0^{+}$, one obtains the non-isothermal Maxwell-Stefan diffusion equations: 
\begin{equation}
\begin{aligned}
& \frac{\partial \rho_{i}}{\partial t} 
  + \frac{\partial }{\partial {\bfx}}( \rho_{i} {\bfu}_i) = 0, 
\\
& \frac{3 m_{0}}{5 m_i} \frac{\partial}{\partial {\bfx}}(\rho_{i} T) 
  = 
  %\frac{1}{\alpha} 
	%\int_{\mathbb{R}^3}\!\! \bfv {Q}_i^{m}(f^\alpha_i, f^\alpha_i)\,d{\bfv} 
	2\pi\sum_{\substack{j=1\\ j\neq i}}^{S}||b_{ij}||_{L^1} \frac{m_j}{m_i+m_j}\rho_i \rho_j ({\bfu}_j-  {\bfu}_i)
\\
& \frac{3}{2} \frac{\partial }{\partial t}\sum_{i=1}^{S} \left( \rho_{i} T \right) 
  + \frac{5}{2} \frac{\partial }{\partial {\bfx}} \sum_{i=1}^{S} \left( \rho_{i} {\bfu}_i T \right) 
  = 0,
\end{aligned}
\label{eq:non-isothermal_MS_equations}
\end{equation}
where
\begin{equation*}
  \rho_{i} = \lim_{\alpha \rightarrow 0^{+}} \rho_{i}^\alpha 
  \quad \text{and} \quad 
  {\bfu}_i = \lim_{\alpha \rightarrow 0^{+}}{\bfu}_i^\alpha.
\end{equation*}
Equations \eqref{eq:Macroscopic_equations} and \eqref{eq:non-isothermal_MS_equations} are up to scaling completely equivalent to the equations derived in \cite{boudin2015maxwell,hutridurga2017maxwell}.

\section{Entropy balance law} 

To derive the macroscopic entropy balance law from the BEs \eqref{eq:system_of_Boltzmann_equations}, we first have to define the entropy flux and the entropy production. 

\begin{df}
Entropy flux $\boldsymbol{\Phi}(t,\bfx)$ and entropy production $D(t,\bfx)$ are defined as:
\begin{gather} 
  \boldsymbol{\Phi}(t,\bfx) := \sum_{i=1}^{S} \boldsymbol{\Phi}_{i}(t,\bfx),  \quad
  \boldsymbol{\Phi}_{i}(t,\bfx) := - k_{\mathrm{B}} 
    \int_{\mathbb{R}^{3}} \bfv f_i \log (b_{i} f_i) d\bfv, 
    \nonumber
   \\ 
   D(t,\bfx) := \sum_{i=1}^{S} D_{i}(t,\bfx), \quad
   D_{i}(t,\bfx) := - k_{\mathrm{B}} \int_{\mathbb{R}^{3}} {Q}_i^{m}(f_i, f_i) \log (b_{i} f_i) d\bfv
    - k_{\mathrm{B}} \sum_{\substack{j=1\\ j\neq i}}^{S} 
    \int_{\mathbb{R}^{3}} {Q}^b_{ij}(f_i, f_j) \log (b_{i} f_i) d\bfv. 
   \nonumber
\end{gather}
\end{df}

By introducing dimensionless quantities:
\begin{equation*}
\hat{\boldsymbol{\Phi}}_{i} = \frac{L^{3}}{k_{\mathrm{B}}N c_{0}}\boldsymbol{\Phi}_{i}, 
  \quad 
  \hat{D}_{i} = \frac{L^{6}}{k_{\mathrm{B}}N^2 4\pi r^2 c_{0}}D_{i},
\end{equation*}
and dropping hat notation for convenience, one may determine the dimensionless partial entropy fluxes and partial entropy productions: 
\begin{equation} \label{eq:species_entropy_production_integral}
\begin{split}
  \boldsymbol{\Phi}_{i}^{\alpha}(t,\bfx) & = - \int_{\mathbb{R}^{3}} \bfv f_i^{\alpha} 
    \log (b_{i} f_i^{\alpha}) d\bfv, 
  \\
  D_{i}^{\alpha}(t,\bfx) & = - \int_{\mathbb{R}^{3}} {Q}_i^{m}(f^\alpha_i, f^\alpha_i) 
    \log (b_{i} f_i^{\alpha}) d\bfv
    - \sum_{\substack{j=1\\ j\neq i}}^{S} 
    \int_{\mathbb{R}^{3}} Q^{b}_{ij}(f^\alpha_i, f^\alpha_j) 
    \log (b_{i} f_i^{\alpha}) d\bfv.
\end{split}
\end{equation}
Setting $\psi_i(\bfv) = - \log (b_{i} f_{i}^{\alpha})$ in \eqref{eq:moment_equations}, one obtains the following scaled entropy balance law:
%Multiplying each Boltzmann equation \eqref{eq:system_of_Boltzmann_equations} with a test function $\psi(\bfv) = - k_{\mathrm{B}} \log (b_{i} f_{i})$, integrating over the velocity space and summing up over all species, one obtains the following entropy balance law:
%\begin{equation}
%  \frac{\partial H}{\partial t} 
%    + \frac{\partial \boldsymbol{\Phi}}{\partial \bfx} = D.
%  \label{eq:entropy_balance_law}
%\end{equation}
%Using dimensionless variables: 
%\begin{equation*}
%  \widehat{\boldsymbol{\Phi}}_{i} = \frac{L^{3}}{k_{\mathrm{B}} N c_{0}} \boldsymbol{\Phi}_{i}, 
%  \quad 
%  \widehat{D}_{i} = \frac{L^{6}}{k_{\mathrm{B}} N^{2} c_{0} \times 4 \pi r^{2}} D_{i}, 
%\end{equation*}
%the entropy balance law \eqref{eq:entropy_balance_law} is transformed into the following dimensionless form (hats are dropped): 
\begin{equation}
  \alpha \frac{\partial H^{\alpha}}{\partial t} 
    + \frac{\partial \boldsymbol{\Phi}^{\alpha}}{\partial \bfx} 
    = \frac{1}{\alpha} D^{\alpha}.
  \label{eq:entropy_balance_law_dimensionless}
\end{equation}
It is a matter of straightforward computation to show that dimensionless entropy densities and fluxes read: 
\begin{equation} \label{eq:species_entropy_density_and_flux}
  H_{i}^{\alpha}(t,\bfx) = \left\{ \log \left[b_i \left(\frac{5}{3m_0}\right)^{\frac{3}{2}} 
    \rho_i^{\alpha} \left( 
    \frac{m_i}{2\pi T} \right)^{\frac{3}{2}}\right] - \frac{3}{2} \right\} \rho_i^{\alpha}, 
  \quad 
  \boldsymbol{\Phi}_{i}^{\alpha}(t,\bfx) = \alpha H_{i}^{\alpha}(t,\bfx) {\bfu}_i^{\alpha}
\end{equation}
\begin{comment}
\begin{equation}
\begin{split}
  H_{i}(t,\bfx) & = -\int_{\mathbb{R}^3}f_i^\alpha\log (b_i f_i^\alpha)\, d{\bfv} 
    = \left\{ \log \left[b_i \left(\frac{5}{3m_0}\right)^{\frac{3}{2}}\rho_i \left( \frac{m_i}{2\pi T} \right)^{\frac{3}{2}}\right] - \frac{3}{2} \right\} \rho_i, 
  \\ 
  \boldsymbol{\Phi}_{i}(t,\bfx) & = -\int_{\mathbb{R}^3}{\bfv}f_i^\alpha\log (b_i f_i^\alpha)\, d{\bfv} 
    = \alpha \left\{ \log \left[b_i\left(\frac{5}{3m_0}\right)^{\frac{3}{2}}\rho_i \left(\frac{m_i}{2\pi T} \right)^{\frac{3}{2}} \right] - \frac{3}{2} \right\} \rho_i {\bfu}_i
\end{split}
\end{equation}
\end{comment}
However, computation of the entropy production is summarized in the following Theorem.
%%%%%%%%%%%%%%%%%%%%%%%%%%%%%%%%%%%%%%%%%%%%%%%%%%%%%%%%%%%%%%%%%%%%%%
\begin{thm} \label{TH:Entropy_production}
The dimensionless partial entropy productions read: 
\begin{equation}
  D_i^{\alpha} = \alpha^{2} \frac{10\pi}{3 T m_0} \sum_{\substack{j=1\\ j\neq i}}^{S}  
    ||b_{ij}||_{L^1}\frac{m_i m_j \rho_i^{\alpha} \rho_j^{\alpha}}{m_i + m_j} 
    \left[ \frac{m_i {\bfu}_i^{\alpha} + m_j {\bfu}_j^{\alpha}}{m_i + m_j} 
    \cdot ({\bfu}_j^{\alpha} - {\bfu}_i^{\alpha}) - {\bfu}_i^{\alpha} 
    \cdot ({\bfu}_j^{\alpha} - {\bfu}_i^{\alpha}) \right],
  \label{eq:species_entropy_production}
\end{equation}
while the dimensionless total entropy production have the form: 
\begin{equation}
 D^{\alpha} = \sum_{i=1}^{S} D_i^{\alpha} = 
    \alpha^{2} \frac{5\pi}{3 T m_0} \sum_{i=1}^{S}\sum_{\substack{j=1\\ j\neq i}}^{S} 
    \left[ ||b_{ij}||_{L^1} \frac{m_im_j}{m_i+m_j} \rho_i^{\alpha} \rho_j^{\alpha} 
    \left| {\bfu}_j^{\alpha} - {\bfu}_i^{\alpha} \right|^2 \right].
  \label{eq:entropy_production}
\end{equation}
\end{thm}
%%%%%%%%%%%%%%%%%%%%%%%%%%%%%%%%%%%%%%%%%%%%%%%%%%%%%%%%%%%%%%%%%%%%%
\begin{pf}
To obtain the partial entropy, we need to evaluate the integrals given in equation \eqref{eq:species_entropy_production_integral} and we will focus on the case Maxwell molecules. This means that the collision cross-sections depend on the ${\bfv}, {\bfv}_*, \boldsymbol{\sigma}$ through the deviation angle $\theta \in [0,\pi]$ between ${\bfv}-{\bfv}_*$ and $\boldsymbol{\sigma}$ only. Specifically,
\begin{equation*}
  B_{ij}({\bfv},{\bfv}_*,\boldsymbol{\sigma})=b_{ij}\bigg(\frac{{\bfv}-{\bfv}_*}{|{\bfv}-{\bfv}_*|} \cdot \boldsymbol{\sigma}\bigg)=b_{ij}(\cos\theta). 
\end{equation*}
Moreover, we assume that $b_{ij}$ is an even function and is such that $b_{ij}\in L^1(-1,1)$ following Grad's angular cut-off assumption. Therefore,
\begin{equation}
  \int_{\mathbb{S}^2}B_{ij}({\bfv},{\bfv}_*,\boldsymbol{\sigma})\,d\boldsymbol{\sigma}
  %=2\pi\int_{-1}^1 b_{ij}(\zeta) d\zeta 
  = 2\pi||b_{ij}||_{L^1}\quad \text{and} \quad \int_{\mathbb{S}^2}\boldsymbol{\sigma}B_{ij}({\bfv},{\bfv}_*,\boldsymbol{\sigma})\,d\boldsymbol{\sigma}
  %=\int_{\mathbb{S}^2}\boldsymbol{\sigma}b_{ij}(\cos\theta)\,d\boldsymbol{\sigma}
  =0,
  \label{eq:integral_over_the_sphere}
\end{equation}
see \cite{boudin2015maxwell}. Furthermore, using the definition of the species distribution functions \eqref{eq:scaled_distribution_function} in $\log (b_if^\alpha_i)$, one can rewrite equation \eqref{eq:species_entropy_production_integral} as follows:
%To compute the integral \eqref{eq:entropy_production_integral}, observe that
%\begin{equation}
%  \log ({\color{blue}b_i}f^\alpha_i)=\log \Bigg[{\color{blue}b_i}\bigg(\frac{5}{3m_0}\bigg)^{\frac{3}{2}}\rho_i\bigg(\frac{m_i}{2\pi T}\bigg)^{\frac{3}{2}}\Bigg]-\frac{5m_i}{6 T m_0}\big(|{\bfv}|^2-2\alpha {\bfv}\cdot{\bfu}_i+\alpha^2 |{\bfu}_i|^2\big).
 % \label{eq:log_f}
%\end{equation}
%Substituting equation \eqref{eq:log_f} into equation \eqref{eq:species_entropy_production_integral} and then expand the resulting equation, we obtain:
\begin{align}
 \nonumber D_i^{\alpha} & = \frac{5m_i}{6 T m_0} k_B 
   \underbrace{\int_{\mathbb{R}^3}|{\bfv}|^2{Q}_i^{m}(f^\alpha_i, f^\alpha_i)\, d{\bfv}}_{\cal A}
  -\alpha \frac{10m_i k_B}{6 Tm_0} {\bfu}_i^{\alpha} \cdot 
  \underbrace{\int_{\mathbb{R}^3} {\bfv}{Q}_i^{m}(f^\alpha_i, f^\alpha_i)\, d{\bfv}}_{\cal B}
  \\
  %&-\Bigg\{{\alpha}^2\frac{5m_i}{6 T m_0}|{\bfu}_i|^2-\log \Bigg[\bigg(\frac{5}{3m_0}\bigg)^{\frac{3}{2}}\rho_i\bigg(\frac{m_i}{2\pi T}\bigg)^{\frac{3}{2}}\Bigg]\Bigg\}\underbrace{\int_{\mathbb{R}^3}{Q}_i^{m}(f^\alpha_i, f^\alpha_i)\, d{\bfv}}_{\cal C}
  %\\
 & + \frac{5m_i}{6 T m_0} k_B \sum_{\substack{j=1\\ j\neq i}}^{S} 
   \underbrace{\int_{\mathbb{R}^3} |{\bfv}|^2 {Q}^b_{ij}(f^\alpha_i, f^\alpha_j) \, 
   	d{\bfv}}_{\cal E} 
   - \alpha \frac{10 m_i k_B}{6 T m_0} {\bfu}_i^{\alpha} \cdot 
   \sum_{\substack{j=1\\ j\neq i}}^{S} 
   \underbrace{\int_{\mathbb{R}^3}{\bfv}{Q}^b_{ij}(f^\alpha_i, f^\alpha_j)\, d{\bfv}}_{\cal F}
  \label{eq:entropy_production_integral_expanded} \\
  &\!+\! k_B \Bigg\{\frac{{\alpha}^2 5m_i}{6 T m_0}|{\bfu}_i^{\alpha}|^2 - \log \Bigg[b_i\bigg(\frac{5}{3m_0}\bigg)^{\!\frac{3}{2}} \rho_i^{\alpha} \bigg(\frac{m_i}{2\pi T}\bigg)^{\!\frac{3}{2}}\Bigg]\Bigg\}\Bigg(\sum_{\substack{j=1\\ j\neq i}}^{S}\underbrace{\int_{\mathbb{R}^3}\!\!{Q}^b_{ij}(f^\alpha_i, f^\alpha_j) d{\bfv}}_{\cal G}+\underbrace{\int_{\mathbb{R}^3}\!\!{Q}_i^{m}(f^\alpha_i, f^\alpha_i) d{\bfv}}_{\cal H}\Bigg).
  \nonumber
\end{align}
The integrals ${\cal A}$, ${\cal B}$, ${\cal G}$ and ${\cal H}$ all vanish, that is
\begin{equation}
  {\cal A}=0 \quad {\cal B}=0, \quad {\cal G}=0 \quad \text{and} \quad{\cal H}=0.
  \label{eq:cal_A_B_C_G}
\end{equation}
To show that the integrals ${\cal H}$ and ${\cal G}$ both vanish, we use the weak formulation of the mono-species and bi-species collision operators given in equations \eqref{eq:weak_form_mono_species_collision_operator} and \eqref{eq:weak_form_bi_species_collision_operator}, respectively, with $\psi({\bfv})=1$. To show that the integrals ${\cal A}$ and ${\cal B}$ both vanish, we use the weak formulation of the mono-species collision operator \eqref{eq:weak_form_mono_species_collision_operator} with $\psi({\bfv})=|{\bfv}|^2$ and $\psi({\bfv})={\bfv}$, respectively, together with the conservation laws of kinetic energy  and linear momentum for mono-species collisions: equations \eqref{eq:conservation_of_momentum_and_kinetic_energy}$_2$ and  \eqref{eq:conservation_of_momentum_and_kinetic_energy}$_1$ with $m_j$ replaced by $m_i$. 
Furthermore, using the weak form of the bi-species collision operator \eqref{eq:weak_form_bi_species_collision_operator} with $\psi({\bfv})=|{\bfv}|^2$ and  $\psi({\bfv})={\bfv}$, respectively, together with equations \eqref{eq:post_collision_velocities} and \eqref{eq:integral_over_the_sphere}, one obtains that the values of the integrals ${\cal E}$ and ${\cal F}$ are as given below:
\begin{equation}
  {\cal E}=\alpha^2 4\pi||b_{ij}||_{L^1} 
  \frac{m_j \rho_i^{\alpha} \rho_j^{\alpha}}{(m_i+m_j)^2}(m_i {\bfu}_i^{\alpha} + m_j{\bfu}_j^{\alpha}) \cdot ({\bfu}_j^{\alpha} - {\bfu}_i^{\alpha}), \quad 
  {\cal F} = \alpha \frac{m_j}{m_i+m_j} 2\pi||b_{ij}||_{L^1} \rho_i^{\alpha} \rho_j^{\alpha} \big( {\bfu}_j^{\alpha} - {\bfu}_i^{\alpha} \big).
  \label{eq:E_F}
\end{equation}
Substituting equations \eqref{eq:cal_A_B_C_G} and \eqref{eq:E_F} into \eqref{eq:entropy_production_integral_expanded} gives \eqref{eq:species_entropy_production}. Summing equation \eqref{eq:species_entropy_production} over all species gives  \eqref{eq:entropy_production}.
\end{pf}

\bibliography{M-S-MEP}

\end{document}